\documentclass[twocolumn,showpacs,preprintnumbers,amsmath,amssymb]{revtex4}
\usepackage{epsfig,graphicx}
\usepackage{dcolumn}
\usepackage{bm}

\begin{document}
\draft

\title{Promote cooperation by localised small-world communication}

\author{Zi-Gang Huang,$^{1}$ Sheng-Jun Wang,$^{1}$ Xin-Jian Xu,$^{2}$ and
Ying-Hai Wang$^{1}$\footnote{For correspondence:yhwang@lzu.edu.cn}}
\address{$^{1}$Institute of Theoretical Physics, Lanzhou University, Lanzhou Gansu 730000, China\\
$^{2}$Departamento de F\'{i}sica da Universidade de Aveiro, 3810-193
Aveiro, Portugal}

\date\today

\begin{abstract}
The emergence and maintenance of cooperation within sizable groups
of unrelated humans offer many challenges for our understanding. We
propose that the humans' capacity of communication, such as how many
and how far away the fellows one can build up mutual communication,
may affect the evolution of cooperation. We study this issue by
means of the public goods game (PGG) with a two-layered network of
contacts. Players obtain payoffs from five-person public goods
interactions on a square lattice (the interaction layer). Also, they
update strategies after communicating with neighbours in learning
layer, where two players build up mutual communication with a power
law probability depending on their spatial distance. Our simulation
results indicate that the evolution of cooperation is indeed
sensitive to how players choose others to communicate with,
including the amount as well as the locations. The tendency of
localised communication is proved to be a new mechanism to promote
cooperation.
\end{abstract}


\maketitle

\section{introduction}

In behavioral sciences, evolutionary biology and more recently in
economics, understanding conditions for the emergence and
maintenance of cooperative behavior among unrelated and selfish
individuals becomes a central
issue~\cite{vonNeumann,MaynardSmith}. In the investigation of this
problem, the most popular framework is game theory together with
its extensions involving evolutionary
context~\cite{Hofbauer1998,Cressman2003}. The public goods game
(PGG), which attracted much attention from economists, is a
general paradigm to explain cooperative behavior through group
interactions~\cite{JHKagel}. The PGG model is characterised by
groups of cooperators doing better than groups of defectors, but
defectors always outperforming the cooperators in their group. In
typical examples, the individual contributions are multiplied by a
factor $r$ and then divided equally among all players. With $r$
smaller than the group size, this is an example of a social
dilemma: every individual player is better off defecting than
cooperating, no matter what the other players do. Groups would
therefore consist of defectors only and forego the public good.

Considerable efforts have been concentrated on exploration of the
origin and persistence of cooperation. During the last decades,
five rules, namely, kin selection~\cite{Hamilton1964}, direct
reciprocity~\cite{Trivers,Axelrod}, indirect
reciprocity~\cite{Alexander,Nowak3}, network (or spatial)
reciprocity~\cite{Nowak1,Ohtsuki2006,Nowak1992,Hauert2,Szabo,HauertScience2002},
and group selection~\cite{Traulsen2006}, have been found to
benefit the evolution of cooperation in biological and ecological
systems as well as within human societies (for a review,
see~\cite{Nowak2006}). In realistic systems, most interactions
among elements are spatially localised, which makes spatial or
graph models more meaningful. Unlike the other four rules, spatial
games (\emph{i.e.}, network reciprocity) can lead to cooperative
behavior in the absence of any strategic
complexity~\cite{Nowak1,Ohtsuki2006,Nowak1992,GSzabo1998}. In
spatial evolutionary PGG, the cooperators can survive by forming
compact clusters, which minimizes the exploitation by
defectors~\cite{Szabo}. Furthermore, Szab\'{o} and Hauert
\emph{et~al.} have recently discussed the effects of compulsory
and voluntary interactions of players in evolutionary PGG, with
the structured populations bound to regular
lattices~\cite{Szabo,Hauert,HauertScience2002} as well as with the
well-mixed
population~\cite{HauertScience2002,Hauert4,Hauert3,Hauert5}. The
factors such as the voluntary participation~\cite{Hauert}, and
small density of population~\cite{Hauert3}, are found to be
capable of boosting cooperation. More recently, Huang
\emph{et~al.} have proposed an extended public goods interaction
model to study the evolution of cooperation in heterogeneous
population, and proved that scale-free networks of contacts can
lead to more competitive cooperation~\cite{Huang2007}.

In real world, people always wish to make decisions based on a
comprehensive knowledge of the pertinent background, such as the
historical performance of each alternative
choice~\cite{Challet1997}, which they may obtain by learning from
people they are regarding. However, the limited eyereach or
capacity of humans actually exists and induces the unperfect
communication. One can merely learn from a small set of people
(corresponding to the so-called ``role models'' in the context of
cultural evolution~\cite{HauertScience2002,Ohtsuki2007}).
Furthermore, these role models rarely can be extended over the
whole population, but localised to the learner's ``vicinity'',
which abstractly implies the people having similar characteristics
as the learner (such as religious background, age, education,
lifestyle, or social class) that in favour of building up mutual
communication.

As a natural extension of those aforementioned factors, an new
intriguing task is to understand how the limitation of
communication capacity of people influences the cooperative
behavior in real world. In this letter, we will study a spatially
extended PGG on two-layered graphs, where one layer especially
depicts the communication among players, with the aim to find out
how the evolution of cooperation depends on the players'
communications.

\section{The model}

In spatial game models, the players occupying the sites of a graph
can follow one of the two pure strategies, cooperation ($C$) or
defection ($D$). There are two types of contacts among players in
the evolutionary process: players collect payoffs from their
neighbours by playing games with
them~\cite{Nowak1,Nowak1992,Hauert2,GSzabo1998,Hauert,Szabo}, and
then update strategies by learning from neighbours. Thus, the
graph occupied by the players can be split into two layers, an
interaction layer and a learning (communication)
layer~\cite{Ifti2004,Ohtsuki2007,Wu2007}. The former one defines
the interaction neighbourhood (IN), \emph{i.e.}, who plays game
with whom. The latter one specifies learning neighbourhood (LN)
for evolutionary updating or, in other words, defines
who-is-the-role-model-of-whom.

In our model, the IN layer where players have public goods
interactions is a $L\times{L}$ square lattice with periodic
boundary conditions. The group size of the interaction neighbours
in the PGG therefore is $N_{I}=5$. The score achieved in PGG
interactions denotes the reproductive success, \emph{i.e.}, the
probability that other players will adopt the player's strategy.
This score is assumed to be determined merely by a single, typical
PGG involving the player and its four nearest interaction
neighbours~\cite{Szabo}. Thus the payoff of one player $i$ is,
\begin{equation}
P(i)=
\begin{cases}
\frac{rn_{c}}{n_{c}+n_{d}}-1  & \text{if } s(i)=C,\\
\frac{rn_{c}}{n_{c}+n_{d}}  & \text{if } s(i)=D,
\end{cases}
\end{equation}
where $n_{c}$, and $n_{d}$ (with $n_{c}+n_{d}=N_{I}=5$) denote,
respectively, the number of participants choosing $C$, and $D$,
and $s(i)$ denotes the strategy of player $i$. The cooperative
investments are normalised to unity and $r$ specifies the
multiplication factor on the public goods.

%fig@@@@@@@@@@@@@@@@@@@@@@@@@@@@@@@@@@@@@@@@
\begin{figure}
\centerline{\resizebox{8cm}{!}{\includegraphics{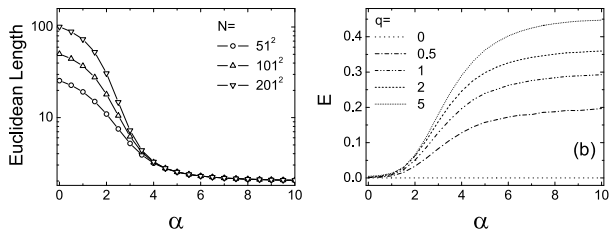}}} \caption{The
average lattice length of the shortcuts $l$ in the LN network with
size $N=L\times{L}=51^{2}$, $101^{2}$, and $201^{2}$ (a). And
clustering coefficient $E$ as a function of $\alpha$ for the LN
with $q=0$, $0.5$, $1$, $2$, and $5$ (b). The plots are average
values over $10$ realizations of the LN network.} \label{figlE}
\end{figure}
%fig@@@@@@@@@@@@@@@@@@@@@@@@@@@@@@@@@@@@@@@@

After each round of the game interactions, for the reference of
the strategy update, each player communicates with its role models
(neighbours in the LN layer), inquiring the individual information
such as the payoff and the strategy. The player's capacity of
communication can be measured by how many and how far away the
role models it can selected. In order to regulate this capacity,
we introduce the LN layer as a variant of the two-dimensional
small-world network~\cite{Kleinberg2000}, in which connections to
further neighbours occur with a tunable power law probability.
This network is constructed by adding shortcuts among the sites on
a $L\times{L}$ square lattice. With periodic boundary condition,
the lattice distance between two sites $(x,y)$ and $(x',y')$ can
be written in a two-dimensional fashion as
\begin{eqnarray}
r_{(x,y),(x',y')}=\Delta{x}+\Delta{y},\nonumber
\end{eqnarray} with
\begin{eqnarray}
&&\Delta{x}=L/2-|(|x-x'|-L/2)|\nonumber\\
&&\Delta{y}=L/2-|(|y-y'|-L/2)|.\nonumber
\end{eqnarray}
This value is actually the length of the shortest path connecting
these two sites through only lattice links. Each site is
additionally linked to $q$ other sites~\cite{intq} by shortcuts
(excluding its original nearest neighbours). Following the idea of
Kleinberg~\cite{Kleinberg2000}, those other sites are selected in
a biased manner: the probability that site $j(x_j,y_j)$ is
selected to be linked to site $i(x_i,y_i)$ by one shortcut depends
on the lattice distance between them in the following way,
\begin{equation}
P(r_{(x_i,y_i),(x_j,y_j)})=\frac{1}{A}r_{(x_i,y_i),(x_j,y_j)}^{-\alpha},\label{Prij}
\end{equation}
where $\alpha$ is a positive exponent and
\begin{equation}
A={\sum_{(x',y')\neq{(x_i,y_i)},(x_i\pm1,y_i),(x_i,y_i\pm1)}r_{(x,y),(x',y')}^{-\alpha}},\label{A}
\end{equation}
is a normalization factor. Obviously, the probability that $i$ and
$j$ are connected is
\begin{equation}
P_{i\leftrightarrow{j}}=1-(1-\frac{r_{(x_i,y_i),(x_j,y_j)}^{-\alpha}}{A})^{2\cdot{q}}.\label{Plink}
\end{equation}

%fig@@@@@@@@@@@@@@@@@@@@@@@@@@@@@@@@@@@@@@@@
\begin{figure*}
\begin{center}
\centerline{\resizebox{11.9cm}{!}{\includegraphics{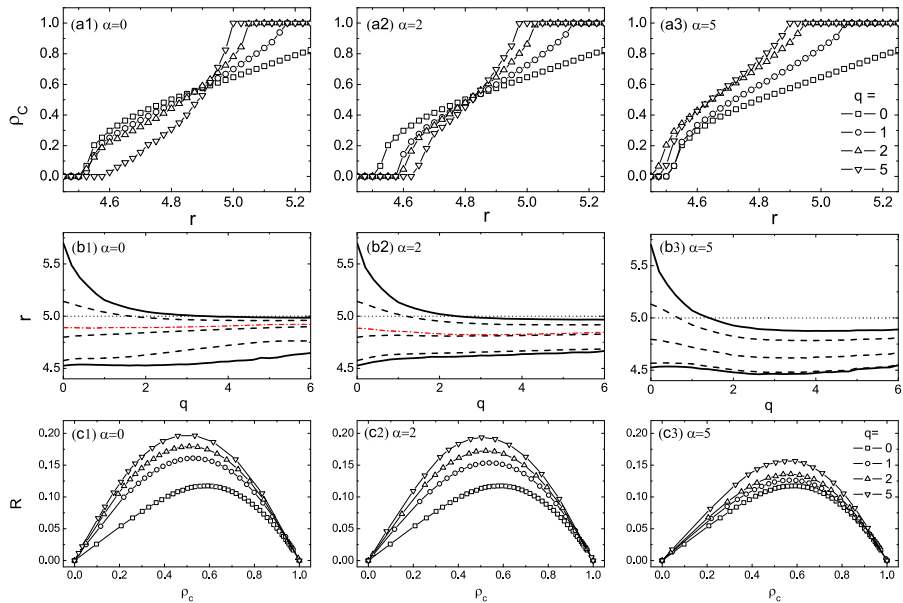}}}
\caption{(a1), (a2), and (a3) show the stationary densities of
cooperators $\rho_{c}$ as a function of multiplication factor $r$
for $q=0$, $1$, $2$, and $5$ denoted by different symbols. (b1),
(b2), and (b3) plot the corresponding phase diagrams. The two bold
solid lines shows the average $r_{c}$ (lower) and $r_{d}$ (upper).
The straight dot line in the middle is $r=5.0$. The three dash
lines from bottom to top present the contours $\rho_{c}=0.25$,
$0.5$, and $0.75$, respectively. The red dash-dot lines in (b1)
and (b2) show the cross points of those $\rho_{c}$ curves of
$q\neq{0}$ with the curves of $q=0$. (c1), (c2), and (c3) show the
strategy update rate $R$ in the dynamical equilibrium state as a
function of $\rho_{c}$. The results are averaged over $10$
realizations of the $N=51^{2}$ systems with $\alpha=0$ (left), $2$
(middle), and $5$ (right).} \label{figPr}
\end{center}
\end{figure*}
%fig@@@@@@@@@@@@@@@@@@@@@@@@@@@@@@@@@@@@@@@@

For this network, when $\alpha=0$, it reduces to the small-world
model with random shortcuts~\cite{Watts1998}. As $\alpha$
increases, one site's shortcuts will be clustered in its vicinity
and two distant sites are less likely to be connected. Here, the
multiple or self-connected edges are not allowed. Such a network
has $qN$ shortcuts and an average degree
$\langle{k}\rangle\approx{4+2q}$. Figure~\ref{figlE} shows the
average lattice length of the shortcuts $l$, and the clustering
coefficient $E$ of this kind graph. We clearly notice the increase
of $E$ with $\alpha$ by reason of the clustered shortcuts.

In this paper, we just consider this kind network as the LN of the
game players where merely the information flow takes place. In
this LN network, the neighbours linked by one edge act as the role
model of each other via mutual communication, and the degree of
each site corresponds to the amount of role models each player
has. For the sake of convenience, we call those role models
introduced by shortcuts the \textbf{additional role models}
(shortly ARMs). Thus, each player has $2q$ ARMs on average. Also,
the biased effect of nonzero $\alpha$, giving rise to the
clustering of shortcuts, can be understood as the ``localization''
of players' communication.

Following previous studies~\cite{GSzabo1998,Szabo}, the evolution
of the present system is governed by random sequential strategy
adoptions, that is, the randomly chosen player $i$ adopts one of
its (randomly chosen) role model $j$'s strategy with a probability
depending on the payoff difference
\begin{eqnarray}
W[s(j)\rightarrow{s(i)}]=\frac{1}{1+exp\{[P(i)-P(j)+\tau]/\kappa\}},\label{Wupdate}
\end{eqnarray}
where $\tau>0$ denotes the cost of strategy change, and $\kappa$
characterises the noise introduced to permit irrational choices.
For $\kappa=0$ the neighbouring strategy $s(j)$ is adopted
deterministically provided the payoff difference exceeds the cost
of strategy change, \emph{i.e.}, $P(j)-P(i)>\tau$. For $\kappa>0$,
strategies performing worse are also adopted with a certain
probability, \emph{e.g.}, due to imperfect information. It is
proved that the dynamics remains unaffected qualitatively when
changing $\kappa$, and $\tau$ within realistic limits. Following
the previous work~\cite{Szabo}, we simply fix the value of
$\kappa$ and $\tau$ to be $0.1$, and concentrate on the general
dynamical properties affected by the structure of LN.

\section{Simulation results}

We study above model by Monte-Carlo (MC) simulations started from
a random initial distribution of C and D strategies. After
appropriate relaxation times, the system can converge to a
dynamical equilibrium state. We characterise this state by the
stationary density of cooperators $\rho_{c}$ averaged over the
last $5000$ MC steps of the $25000$ total sampling steps.
Figures~\ref{figPr} (a1), (a2), and (a3) show the dependence of
stationary density of cooperators $\rho_{c}$ on the multiplication
factor $r$ for the systems with $\alpha=0$, $2$, and $5$,
respectively. The simulation data result from an average over
either ten realizations of independent initial strategy
distributions or ten realizations of the LN networks. For $q=0$,
\emph{i.e.}, when the IN and LN are identical, we recover the
results of ref.~\cite{Szabo}: below the threshold value
$r<r_{c}=4.526(1)$ cooperators quickly vanish (the absorbing
homogenous state with all defectors), whereas for high $r>r_{d}$
defectors go extinct (the absorbing homogenous state with all
cooperators); for intermediate $r$, strategies C and D coexist in
dynamical equilibrium. The $r_{c}$ ($r_{d}$) indicates the value
of $r$ where cooperators (defectors) vanish. In the case that
$q\neq{0}$, \emph{i.e.}, when each player can communicate with
more ARMs via shortcuts, the quantitative properties of the
stationary density $\rho_{c}$ are different [see the curves with
different $q$ in figs.~\ref{figPr} (a1), (a2) and (a3)].

One can find from figs.~\ref{figPr} (a1), and (a2) that, each of
the $\rho_{c}$ curve with $q\neq0$ has a cross point (denoted by
$r_{cross}$) with the curve of $q=0$. The density of cooperators
$\rho_{c}$ for the system with larger $q$ is comparatively smaller
at $r<r_{cross}$ region, but larger $\rho_{c}$ at $r>r_{cross}$
region. That is to say, for the systems with $\alpha=0$ and $2$,
the more available role models will favour cooperators when $r$ is
large, and favour defectors in contrast when $r$ is small.
However, the system with $\alpha=5$ [figs.~\ref{figPr} (a3)] is
obviously different from that of $\alpha=0$ and $2$, that is, the
ARMs can favour cooperators in the whole range of $r$ with respect
to the $q=0$ system, moreover, defectors vanish at much lower $r$.

We plot the phase diagrams of the corresponding systems in
figs.~\ref{figPr} (b1), (b2), and (b3), respectively. The two bold
solid curves respectively show the $r_{c}$ and $r_{d}$ averaged
over $10$ realizations, which divide the region into absorbing
homogenous states of C (upper) and D (lower), as well as the
coexistence regime of the two strategies (intermediate). That is,
for fixed number of ARMs, when varying the factor $r$, two phase
transitions occur between the coexistence region and homogeneous
states of C or D. In the coexistence regime, the three dash lines
present the contours $\rho_{c}=0.25$, $0.5$, and $0.75$,
respectively. The straight dot line in the middle corresponds to
$r=5.0$, larger than which the social dilemma raised by the PGG
with $N_{I}=5$ is relaxed in the sense that each unity investment
has a positive net return. As $q=0$ defectors may exist in the
system and exploit cooperators until $r$ is very larger. However,
from the rapid decline of $r_{d}$ with increasing $q$ we know
that, with the aid of the ARMs, defectors vanish at much smaller
$r$. More, interestingly, even if $r<5.0$, defectors can be
eliminated as long as $q>3.49$ for $\alpha=0$, $q>2.42$ for
$\alpha=2$, or $q>1.41$ for $\alpha=5$. Also, the region of
homogenous C becomes wider as the value of $\alpha$ increases.

The cross points $r_{cross}$ of those $\rho_{c}$ curves of
$q\neq{0}$ with the curve of $q=0$ in fig.~\ref{figPr} (a1) [and
(a2)] are marked by the (red) dash-dot line in the phase diagram.
The coexistence regime of C and D thus is separated into two
regions by this line, with ARMs favouring cooperation in the upper
region, but favouring defection in the lower region. It can be
seen that, the cross points $r_{cross}$ is not sensitive to the
parameter $q$. For the system with $\alpha=2$ the upper region
favouring cooperation is comparatively wider than that with
$\alpha=0$. Moreover, for the system with $\alpha=5$ [see
figs.~\ref{figPr} (a3) and (b3)], those $\rho_{c}$ curves are not
intersectant, and the ARMs favour cooperators in the whole range
of $r$. The curves of $r_{c}$, $r_{d}$, and other contours of
$\rho_{c}$ for various $\alpha$ systems are proved to
asymptotically approach $r=5.0$ with increasing $q$, and finally
collapse into one there ($r=5.0$) for $q\rightarrow{\infty}$,
which implies that each player can learn from the whole
population. That is to say, as a result of the dynamical process
the system with $q\rightarrow{\infty}$ will end up in the
absorbing D state for any $r$ smaller than $5.0$, where the
first-order phase transition from D to C takes place.

%fig@@@@@@@@@@@@@@@@@@@@@@@@@@@@@@@@@@@@@@@@
\begin{figure}
\centerline{\resizebox{7.5cm}{!}{\includegraphics{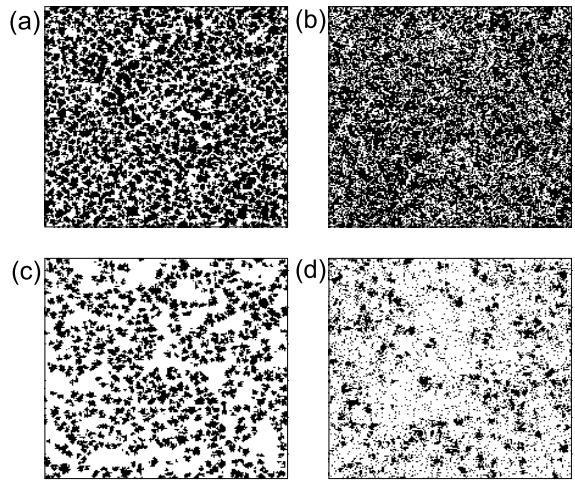}}}
\caption{Snapshots of typical distributions of cooperators (black)
and defectors (white) on square lattice of $N=201^{2}$ for
multiplication factor $r=4.9$ (upper) and $4.6$ (lower). (a) and
(c) correspond to the case with identical IN and LN, while (b) and
(d) correspond to the case of LN with $\alpha=2$ and $q=2$.}
\label{figTu}
\end{figure}
%fig@@@@@@@@@@@@@@@@@@@@@@@@@@@@@@@@@@@@@@@@

We also notice that the ARMs may affect the severity of the
competition between C and D, which is characterised by the rate of
the effective strategy update $R$ (the fraction of players who
adopt the opposite strategy in each generation averaged over
$1000$ MC steps). For the sake of comparison, the rate $R$ are
plotted as a function of $\rho_{c}$ in figs.~\ref{figPr} (c1),
(c2), and (c3) for the systems with $\alpha=0$, $2$, and $5$,
respectively. For a given stationary density $\rho_{c}$, the
system with larger $q$ results in larger rate $R$, which implies
that more ARMs induce more intense competition. This is due to the
additional mutual contacts between C and D, or, in other words the
larger surface between C and D in the LN network, induced by the
additional shortcuts. Also, $R$ exhibits a bell-like form, because
the surface between C and D will shrink with the density
difference of the two strategies. Additionally, for a given value
of $q$, the system with larger $\alpha$ are found to result in
smaller rate $R$.

Figure~\ref{figTu} shows the snapshots of the system with identical
IN and LN (left hand), as well as the system with $q=2$ and
$\alpha=2$ LN (right hand), respectively. For the case of identical
IN and LN [fig.~\ref{figTu}(a)(c)], it is known from PGG and other
cooperation games that, cooperators persist through forming compact
C clusters and thereby reducing exploitation by
defectors~\cite{GSzabo1998,Nowak1,Szabo,Huang2007,Gardens2007}.
However, for the availability of ARMs [fig.~\ref{figTu}(b)(d)], the
C clusters break into smaller pieces. Furthermore, one interesting
phenomenon distinctly exhibited in fig.~\ref{figTu} (d) is, some
rare cooperators may take place at the sites totally surrounded by
defectors, where it is extremely difficult for C to survive because
of the very low payoffs. These sites are found to be connected by
shortcuts to the well-paid C role models who are located in the C
clusters. Therefore, we can say that, the C strategy is adopted
there when players blindly imitate the successful role models
without regard to their own actual `habitats'. This may correspond
to the phenomenon in society that, some people follow like sheep to
the manner of others ignoring the question whether it fits into
their own social surrounding. Interestingly, this unreasonable
behavior simply introduces mutations to the population of D. We
therefore suggest that, it may provide one mechanism to favour
cooperation, for the invasions of C may form mutual protection and
thus survive in the population of D, although the rare C therein are
unstable and hard to propagate by themselves. In addition, these
sites exhibit one path to put up the fluctuating individuals
\cite{Gardens2007}, who alternatively spend some time as cooperators
and some time as defectors.

\begin{figure}
\centerline{\resizebox{8cm}{!}{\includegraphics{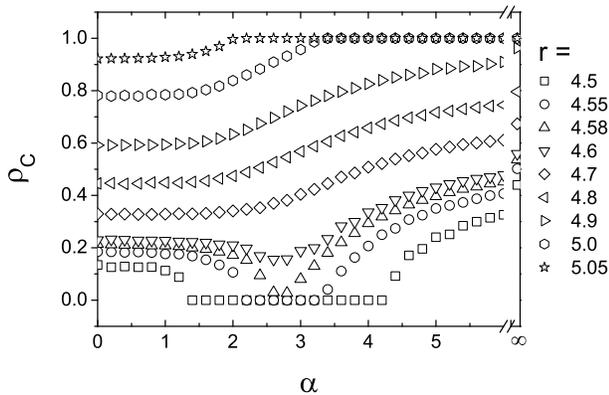}}}
\caption{Stationary densities of cooperators $\rho_{c}$ as a
function of $\alpha$ with $r$ ranges from $4.5$ to $5.05$. The
results are averaged over $10$ realizations of the system with
$N=201^{2}$ and $q=2$.} \label{figPa}
\end{figure}

Finally, we focus our attention on how the localised communication
affects the evolution of cooperation. The stationary density of
cooperators $\rho_{c}$ at different values of $r$ are plotted as a
function of $\alpha$ in fig.~\ref{figPa}. We can see that the
values of $\rho_{c}$ keep almost invariable in the region
$0\leq\alpha<1$. This can be naturally understood from the results
in fig.~\ref{figlE}, as well as the analysis in the
ref.~\cite{Sen20012002,Chatterjee2006,Shang2006} that, in this
region the LN network behaves as a small world with the
topological properties changing slowly with $\alpha$. However, in
the large $\alpha$ region, which implies that the LN is distinctly
localised, cooperation is favoured more compared to the case with
$\alpha=0$. We have known that, the cooperators protect each other
and survive by forming clusters. Thus, when the shortcuts are
clustered by larger $\alpha$, one cooperator is more likely to
select role models within its own C cluster via shortcuts. That is
to say, the mutual protection among cooperators is enhanced by
large $\alpha$, and thereby cooperation is favoured. This
mechanism is obviously reflected by the monotonic increase of
$\rho_{c}$ with $\alpha$ at large $\alpha$ region. Furthermore,
the best communication strategy for players to promote cooperation
is $\alpha\rightarrow\infty$, which corresponds to the extremely
localised selection of ARMs, \emph{i.e.}, the LN network with the
shortcuts merely connects the next-nearest neighbours through
lattice links.

%fig@@@@@@@@@@@@@@@@@@@@@@@@@@@@@@@@@@@@@@@@
%fig@@@@@@@@@@@@@@@@@@@@@@@@@@@@@@@@@@@@@@@@

It is also notable in fig.~\ref{figPa} that, the $\rho_{c}$
monotonically increases with $\alpha$ when $r$ is comparatively
large (about $r>4.6$), however, the nonmonotonic behaviors of
$\rho_{c}$ occur for the cases of smaller $r$. This result is
proved to remain unaffected qualitatively for different vales of
$q$ within realistic limits. As it is well known, the cooperators
located along the boundary of the C clusters (the so-called
\textbf{boundary C} in the following text) outweigh the losses
against defectors by gains from interactions within the C
cluster~\cite{GSzabo1998,Hauert2}. These boundary C gain low
payoffs and thus would be sensitive to the introduction of the
ARMs, \emph{i.e.}, what kind of role models they meet with via
shortcuts really matters. The system at the coexistence state is
composed of cooperators as well as two kinds of defectors, the
defector located right at the boundary of C cluster which we named
D$_1$, and the defector surrounded by the same-strategy
interaction neighbours which we named D$_2$. One D$_1$ would
accumulate comparatively higher payoff by exploiting the boundary
C, and then act as an attractive role model of its learning
neighbours. However, one D$_2$ gains zero payoff, and thus will
not affect the stability of others' strategy. They put up
fluctuating individuals if they are linked to well-paid C by
shortcuts, as that discussed above. In our model, the players will
choose ARMs nearer and nearer to its vicinity as $\alpha$
increases from $0$. When $\alpha$ is around $2.5$, the lattice
lengths of these shortcuts (see fig.~\ref{figlE}) approximately
reach the general size of the C clusters, then the probability for
the boundary C to meet with the well-paid D$_1$ players becomes
larger, which would pose a high risk to the stability of the
boundary C. The C cluster becomes constricted and thus cooperation
is depressed when the boundary C imitates D$_1$. This effect can
be clearly observed from the comparatively smaller $\rho_{c}$
around $\alpha=2.5$ at small $r$ (see fig.~\ref{figPa}). However,
when the value of $r$ is large, this effect is weakened by the C
clusters highly crowded [see fig.~\ref{figTu}(b)].

\section{Conclusion}

In summary, we have shown how the communication among players may
affect the evolution of cooperation by means of a two-layered PGG
model. The learning layer is constructed as a Kleinberg
small-world network, where two players can communicate with
probability depending on their spatial Euclidean lattice distance
in the power-law form controlled by an exponent $\alpha$. The
players' capacity of communication is characterised by the number
and the distance of the role models from whom they can learn the
individual information. The biased effect of nonzero $\alpha$,
which gives rise to the preferential selection of role models near
the vicinity, corresponds to the localization of players'
communication induced by limited capacity. Our simulation results
indicate that, the communication among players plays a highly
important role in the evolution of cooperation: the density of C
is crucially influenced by the number ($q$) and the location of
the ARMs; the coexistence region of C and D is reduced and $r_{c}$
($r_{d}$) is expected to tend to $5$ (the group size $N_{I}$) if
$q$ increases; in addition, for certain density of C, more
available ARMs result in more intense competition between C and D.
Moreover, the communication via shortcuts (\emph{i.e.} the ARMs)
are found to introduce mutation to the population of D, which
might be helpful to the emergence of C. It is also notable that,
the localised communication with large $\alpha$ can favour
cooperators.

Our model gives a crude simulation of real social behavior.
However, it does catch a few features of potential interest. The
most interesting feature is the localised communication favouring
cooperation. We suggest that, the limitation of humans' capacity
to build up communication system, which results in the localised
selection of role models, would be a new mechanism supporting the
emergence and persistence of cooperation in society. The other
feature is that a population of cheaters would be invaded by
cooperators by the presence of long-range social connections.

\acknowledgments

We thank Dr. Zhi-Xi Wu for helpful discussion. This work was
supported by the Natural Science Foundation of China
(No.~10775060). X.-J. Xu acknowledges financial support from FCT
(Portugal), Grant No. SFRH/BPD/30425/2006.


\begin{references}
\bibitem{intq}
When $q$ is not an integer, each node sends out $int(q)$ shortcuts
deterministically, as well as one more shortcut with probability
$q-int(q)$. All these shortcuts are generated independently with the
probability Eq. (\ref{Prij}).
\end{references}
\end{document}